\documentclass[12pt]{article} 
\usepackage{graphicx}
\UseRawInputEncoding
	
\title{A large scale pattern from optical quasar polarization vectors}  
\author{{\it Richard Shurtleff~}\thanks{affiliation and mailing 
address: Department of Science, 
Wentworth Institute of Technology, 550 Huntington Avenue, 
Boston, MA,  02115, USA, e-mail: shurtleffr@wit.edu, momentummatrix@yahoo.com}} 
\begin{document}

\maketitle 

\begin{abstract} 


The 355 optically polarized QSOs have redshifts from $0.061$ to $3.94\,$ and are spread out over the sky except for a $60^{\circ}$ band centered on the Galactic Equator. The data we analyze was measured, collected and published by others. Here, we apply tests suitable for large-scale samples and find that the polarization directions align with a significance of $p$ = $1\%$ by one test and are correlated by a second test with  $p$ = $5\%,$ each $p$-value uncertain within a factor of about $2$. The tests return a preferred Cartesian coordinate system that fortuitously aligns well with the Milky Way Galaxy with a significance of  $p$ = $3\%.$ Thus, the Hub Tests' results combined together imply the polarization directions are correlated with a significance that is much less than $1\%,$ which is remarkable for such a large-scale sample. 


\vspace{0.5cm}
Keywords: Quasars: general ; Polarization ; Large scale structure 
 
\vspace{0.5cm}


\end{abstract}
\pagebreak

\section{Introduction} \label{intro}

	Many recent studies of astronomical sources involve large-scale samples and have uncovered large scale correlations. For examples, a dipole has been found in the angular distribution of quasars (QSOs),  the low-order harmonics of the Cosmic Microwave Background reveal structure, and an asymmetry in the distribution of spiral galaxy spin directions  has been determined.\cite{Secrest_2021,Tegmark_2003,Shamir, Longo} For a review of the many extreme-scale observations that may violate the Cosmological Principle consult  \cite{Aluri2022}. See Section A of Part V in Ref. \cite{Aluri2022} for QSO polarization alignments. Besides observations, there are theoretical ideas associated with the evolution of the universe that include whole-sky scale effects.\cite{Araujo_2015} 	
	
	Partial linear polarization is a property of the electromagnetic radiation received from some astronomical objects and utilized, for example, with mapping magnetic fields.\cite{PolarizationToMagFields} In this article, we look for correlations of the polarization directions of a large-scale sample of optical QSO sources.
	
	
	A 355 partially polarized optical QSO sample is presented and described in Refs. \cite{Hutsemekers2005, 355QSOcatalog}. The catalog is from published literature and is available online. The data is re-analyzed here. The originating articles, including Ref. \cite{Hutsemekers2005} and references therein, describe the source selection, measurement methods and data reduction in detail.  Here, the data is analyzed with the Hub Tests of Alignment and Avoidance. A supplemental Mathematica \cite{Wolfram} notebook verifies the calculations in this article.\cite{MMAnotebook} Both Hub Tests are described briefly in the Appendix. 
	
	The 355 QSOs are scattered over the Northern and Southern Galactic hemispheres, $195$ in the North and $160$ in the South. None of the sources are within $30^{\circ}$ of the Galactic Equator. With a sample that is so widely distributed on the sky, it is important to use correlation tests that are appropriate for large-scale samples. 
	
	
	The Hub Tests of Alignment and Avoidance look for correlations of the type exhibited by Local North and Local East vectors. The well-known concepts of Local North and Local East apply globally, so the tests are suitable for large-scale samples. The Hub Tests allow any point $H$ on the sky, except points that happen to coincide with sources, to serve as a virtual North Pole. For each $H,$ one asks how well the polarization directions match Local North or Local East vectors. For details, see the Appendix.
	
	Other alignment tests are applied to this sample in the original article by Hutsem\'{e}kers {\textit{et al.}}, the article that presents the catalog, Ref.  \cite{Hutsemekers2005}. Those tests, notably the S and Z-tests and similar tests, compare the polarization directions directly.\cite{Jain2004, Contigiani2017} Such tests are complicated by the geometry of curved surfaces: comparing the directions of vectors at different points on a curved surface depends on establishing a consistent parallel transport rule and deciding where to transport the vectors for comparison. Not all issues can be resolved.  For a sample of three or more vectors, it is straightforward to show that the angular separations of transported vectors depends on the choice of the preferred reference point where the vectors are transported to for comparison. The issues have been faced and somewhat mitigated.\cite{Jain2004, Contigiani2017} 
	
		An earlier analysis of the 355 QSO catalog data using the Hub Tests exists, call it Version~1.\cite{Shurtleff 2013} In Version 1, a simplifying assumption approximates the function mapped in Fig. \ref{MapOfEtaBar} with a quadrupole. Overall, the quadrupole is an excellent fit. However, the approximation moves the location of the alignment function minimum, $H_{\mathrm{align}},$ some $20^{\circ}$ from the actual minimum. That rotates the preferred coordinate system in Version 1 by the same $20^{\circ}\, .$ 
	
	In this article, the function is not approximated. Furthermore, the Hub Tests are now described with a new approach using different concepts. And technological advances in computer equipment and software allow the statistics and uncertainty discussions here to be superior to those sections in Version 1. In Version 1 there were hundreds of random runs; here, there are tens of thousands. 
	

\begin{figure}[ht]  
\centering
\vspace{0cm}
\hspace{0in}\includegraphics[bb=130 10 280 185, height=2.5in,keepaspectratio]{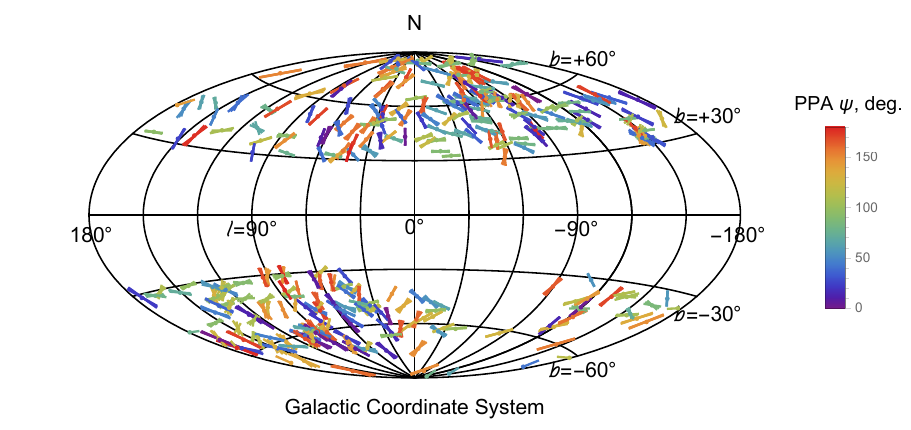}
\caption{(Color online) {\it{The Data, displayed on an Aitoff projection of the whole sky.}}  The 355 QSOs and their polarization directions with their uncertainties are plotted in J2000.0 galactic coordinates. The band of white space reflects the absence of sources within $30^{\circ}$  of the Galactic Equator. The shading color indicates the polarization position angle PPA $\psi$ between the polarization direction of each source and the Local North (N). Both $\psi$ = $0$ and $\psi$ = $180^{\circ}$ are along the Local North/South meridian, so values of $\psi$ loop by joining $\psi$ = $180^{\circ}$ with $\psi$ = $0.$ }
\label{PineNeedleData}
\end{figure}

		
	Section \ref{355QSOs} presents the data and describes the analysis and results. The Hub Test of Alignment finds that the alignment is very significant, with one in every $108 \{_{-78}^{+83}$ randomly directed samples expected to be better aligned, so the significance $p$ is roughly $1\%.$ The Hub Test of avoidance produces a result better than all but one in $19\{_{-12}^{+8}$ random runs, with  $p \approx$ $5\%,$ roughly. 
	
	
	Section \ref{3DFrame} discusses some of the geometrical  consequences of the Hub Test results. The locations of the alignment and avoidance hubs determine a preferred Cartesian coordinate system that is a product of the observed polarization directions of the 355 QSO sources. A determination of the significance of the preferred reference frame orientation shows that $p $ = $3.0\% \pm 0.4\%.$ Fewer than one in about 30 randomly directed samples would be located where the observed data puts the preferred coordinate system.
	
	Finally, Sec. \ref{Discussion} has a summary and brief discussion to end the article. The combination of unlikely outcomes is more unlikely than any one. The combined significance of the Hub Tests' results is remarkable for such a large-scale sample.


\section{355 QSOs} \label{355QSOs}

	The sample consists of $355$ partially linearly polarized optical QSOs.\cite{Hutsemekers2005} The online catalog includes each QSO's position $S_i$ and polarization position angle $\psi_i,$ $i \in $ $\{1, ..., 355\}.$ That data is plotted in Fig. \ref{PineNeedleData}.  Cutoffs to data require that the position angles $\psi$ must have uncertainties $\sigma \psi_i$  less than $14^{\circ}$ and the polarization degree must be at least $0.6\%.$ Redshifts range from $0.61$ to $3.94 \,$. For details about the data consult Ref. \cite{Hutsemekers2005} and the originating references listed in the catalog.
	
	
	It turns out to be fortuitous that no sources are within $30^{\circ}$ of the Galactic Equator. This orients the large-scale sample with respect to the Galaxy. 
	
	\begin{figure}[ht]  
\centering
\vspace{0cm}
\hspace{0in}\includegraphics[bb=130 10 280 185, height=2.5in,keepaspectratio]{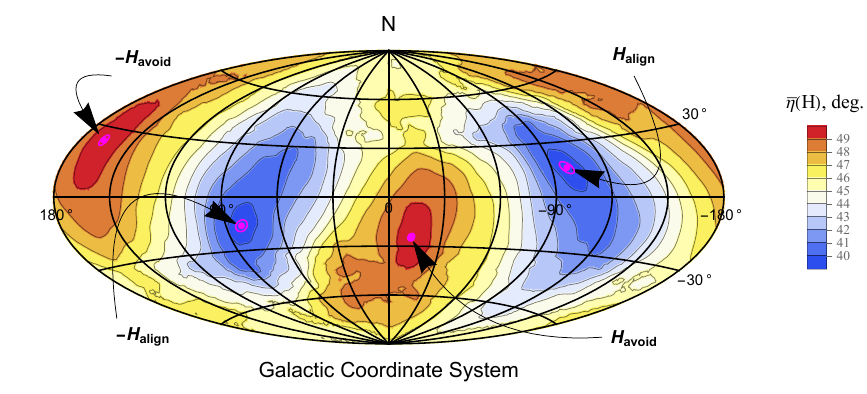}
\caption{(Color online) {\it{The Average Alignment Angle Function $\bar{\eta }(H),$ Eq. \ref{Eq1}.}} Based on the alignment function $\bar{\eta }(H),$ the Hub Tests of Alignment and Avoidance produce four quantities to determine correlations among the samples' polarization directions.  Two of the quantities are the function's extreme values, $\bar{\eta }_{\mathrm{min}}$ and $\bar{\eta }_{\mathrm{max}}.$ And there are the locations where the two extremes occur. These are called `hubs'. Where $\bar{\eta }(H)$ is a minimum, we have the alignment hubs  $\pm H_{\mathrm{Align}}.$ The two diametrically opposed locations where the maximum $\bar{\eta }_{\mathrm{max}}$ is attained are called the avoidance hubs $\pm H_{\mathrm{Avoid}}.$}
\label{MapOfEtaBar}
\end{figure}

	Since the partially polarized light is emitted by QSOs at redshifts at or beyond $z \geq$ $0.061$ and the light is detected here, the light emitted by the QSOs travels through the Galaxy along paths well approximated by near-field galactic stars.  Based on starlight data, as shown in \cite{Hutsemekers2005},  the QSOs' polarization directions are not significantly impacted by local interaction with the Milky Way. Thus, we accept the conclusion in Ref. \cite{Hutsemekers2005} that the observed polarization directions are largely unaffected by any interaction of the QSO light with the matter in the Milky Way.
		
	Figure \ref{MapOfEtaBar} shows the fundamental Hub Test function $\bar{\eta }(H).$ The Hub Test gives every point $H$ on the sky the chance to be a Virtual North Pole. That creates a virtual Local North vector at each source $S_{i}$ for each point $H$ on the sky.  The polarization direction at source $S_{i}$ makes an `alignment' angle $\eta_{iH}$ with the virtual Local North at $S_{i}.$ Thus the alignment angle $\eta_{iH}$ is a kind of virtual polarization position angle with respect to the virtual North Pole at $H.$ See Fig. \ref{etaH}.
	
	The average of the $N$ = $355$ alignment angles $\eta_{iH}$ is the Hub Test function $\bar{\eta }(H).$  We have 
	\begin{equation} \label{Eq1}
	\bar{\eta }\left( H \right)  =  \frac{1}{N} \sum _{i=1}^N  \eta _{iH} \; .
\end{equation}
The function is defined at any point that is not a source, $H \neq$ $S_{i}.$  Since polarization vectors are bi-directional, the $\eta_{iH}$ and, consequently, $\bar{\eta }(H)$ can be acute. For more details, see the Appendix, where Eq. \ref{Eq1} is Eq. \ref{EqA1}.

	Small values of $\bar{\eta }\left( H \right)$ in Fig. \ref{MapOfEtaBar} locate places where the polarization directions at the $355$ sources tend to align with the virtual meridians of the Virtual North Pole at $H.$ Large values mean the polarization directions tend to align with the virtual parallels of the Virtual North Pole at $H.$

	The smallest value of $\bar{\eta }\left( H \right),$ call it $\bar{\eta }_{\mathrm{min}},$ occurs at some point $H$ on the sky, call it the `hub' $H_{\mathrm{Align}}.$ With a Virtual North Pole at  $H_{\mathrm{Align}}$ the polarization directions act best as Local North directions. The largest value of $\bar{\eta }\left( H \right),$ called  $\bar{\eta }_{\mathrm{max}},$ occurs at another `hub', $H_{\mathrm{Avoid}}.$ With a Virtual North Pole at $H_{\mathrm{Avoid}},$ the polarization directions act as Local East directions better than putting the Pole at any other point on the sky. 
	
	The small loops around the hubs in Fig. \ref{MapOfEtaBar} are uncertainty ellipses due to the experimental uncertainty in the polarization directions $\psi.$ The contour lines shown on the map would be similarly blurred into bands a couple of degrees wide. The uncertainty of the contour lines is not displayed in the figure.

         \vspace{0.5cm}

	\begin{figure}[ht]  
\centering
\vspace{0cm}
\hspace{0in}\includegraphics[height=2.75in,keepaspectratio]{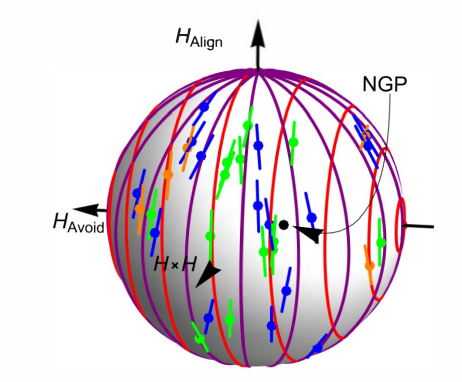}
\caption{(Color online) {\it{Selected data plotted in the preferred coordinate system. }} To avoid clutter, we draw only the sources that have perfect polarization directions, within experimental uncertainty, $\pm \sigma \psi.$ Those directed perfectly along the meridians for a Virtual North Pole at  $H_{\mathrm{Align}}$ are shaded in blue. Those directed along parallels for a Virtual North Pole at  $H_{\mathrm{Avoid}}$ are shaded in orange. Doubly perfect sources are in green.  Meridians for a Virtual North Pole at  $H_{\mathrm{Align}}$ and parallels for a Virtual North Pole at  $H_{\mathrm{Avoid}}$ are drawn in purple and red, respectively, to help guide the eye. 
}
\label{GlobeFromNGP}
\end{figure}

		The Hub Test of Alignment asks where one should place a Virtual North Pole so that the polarization directions best align with meridians. The best examples of this are polarization directions that are parallel to the local meridian within experimental error. Those sources, and the sources with perfect avoidance within experimental error, are plotted in Fig. \ref{GlobeFromNGP}, on a sphere that is rotated to align with the preferred reference system.
	
\begin{figure}[ht]  
\centering
\vspace{0cm}
\hspace{0in}\includegraphics[bb=130 10 280 185, height=2.3in,keepaspectratio]{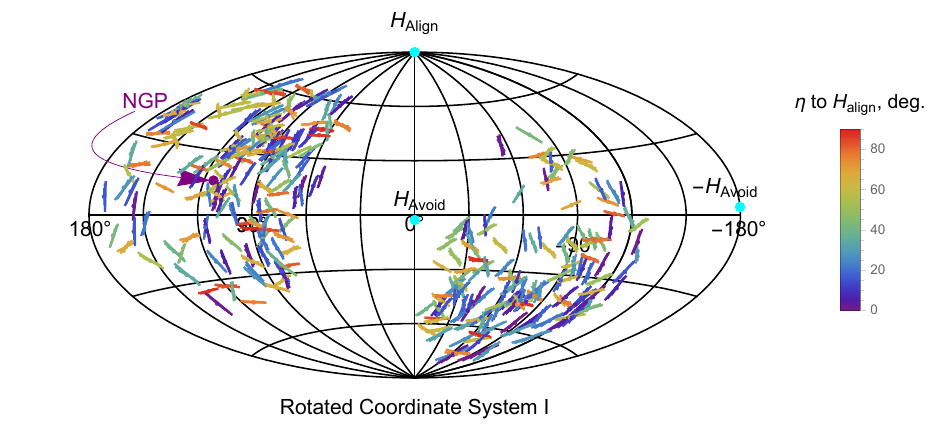}
\caption{(Color online) {{\textit{The polarization vectors as Local Norths.}} A rotation takes the alignment hub $H_{\mathrm{Align}}$ to the North Pole and the avoidance hub $H_{\mathrm{Avoid}}$ to longitude $0^{\circ}.$ The polarization directions act as Local Norths when they are parallel with nearby meridians. Many are. The vectors that best approximate Local Norths, those shaded purple and blue, are widespread, so the alignment is an extreme-scale phenomenon. }}
\label{NisHalign}
\end{figure}

	All sources are plotted in Fig. \ref{NisHalign}, which is rotated to put the alignment hub $H_{\mathrm{Align}}$ at the (virtual) North Pole. One sees that many of the polarization directions align with the meridians and approximate Local North vectors quite well.

		\begin{figure}[ht]  
\centering
\vspace{0cm}
\hspace{0in}\includegraphics[bb=130 10 280 185, height=2.3in,keepaspectratio]{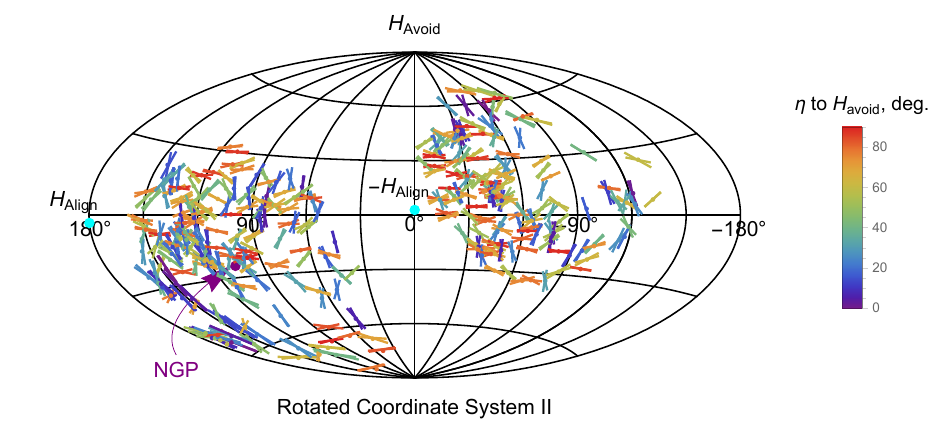}
\caption{(Color online) {{\textit{The polarization vectors as Local Easts.}} Rotating the coordinate system to put the avoidance hub $H_{\mathrm{Avoid}}$ at the North Pole, the plot shows that many of the polarization vectors point along the Local East-West direction. The best examples are those shaded red and orange.  }}
\label{NisHavoid}
\end{figure}

	The Hub Test of avoidance asks how well the polarization directions act as Local East directions when the coordinate system is rotated to place the hubs  $H_{\mathrm{Avoid}}$ and $-H_{\mathrm{Avoid}}$ at virtual poles.  By inspection of Fig. \ref{NisHavoid}, many of the polarizations are directed along parallels, circles of constant latitude.  However, one should also notice that many polarizations follow meridians, making a somewhat bimodal pattern. Consult Ref. \cite{MMAnotebook} for details.

\begin{figure}[ht]  
\centering
\vspace{0cm}
\hspace{0in}\includegraphics[bb=130 10 280 185, height=2.3in,keepaspectratio]{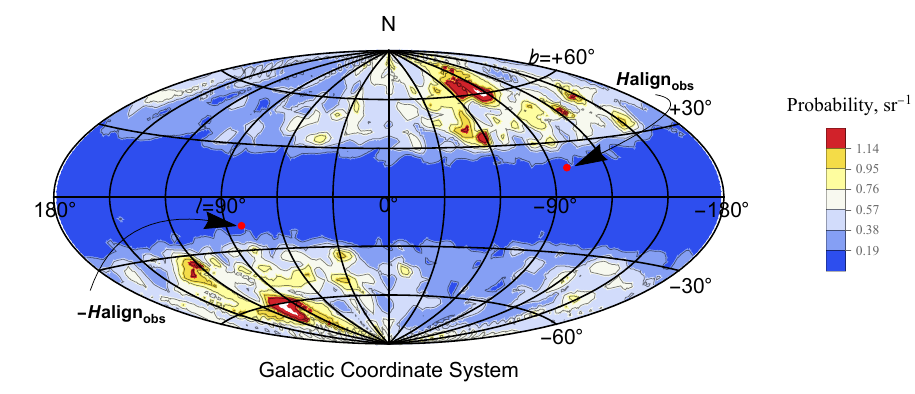}
\caption{(Color online) {{\textit{Probability density of  $25\,000$ random run $H_{\mathrm{Align}}$ hub locations.  } For each random run, the sources' polarization directions are random, which leaves only the locations of  the sources as the remaining relevant property. Since the sources are located at least $30^{\circ}$ away from the Galactic Equator, the figure shows that alignment hubs for randomly directed samples are most likely to occur near the sources. The observed polarization direction put the alignment hubs within $30^{\circ}$ of the Galactic Equator, where random run hubs are unlikely. }  
}}
\label{ProbAlign}
\end{figure}

\begin{figure}[ht]  
\centering
\vspace{0cm}
\hspace{0in}\includegraphics[bb=130 10 280 185, height=2.3in,keepaspectratio]{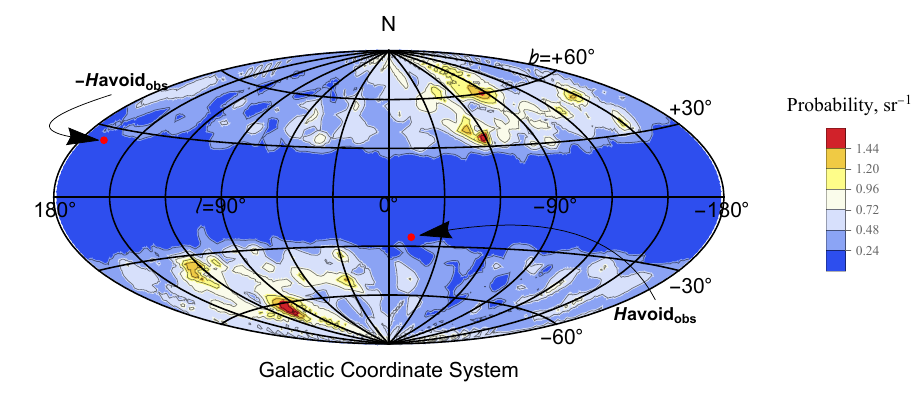}
\caption{(Color online) {{\textit{Probability density of $25\,000$ random run $H_{\mathrm{Avoid}}$ hub locations. } As in Fig. \ref{ProbAlign}, when the QSOs host random directions, avoidance hubs are more likely found near the sources. The patterns evident in the two figures suggest that it requires collective concerted agreement of the polarization directions of a sizable subset of the sources to put a hub far from the sources.  Such agreement is rare for randomly directed sources.  }  
}}
\label{ProbAvoid}
\end{figure}

	         \vspace{0.8cm}

	{\textbf{Uncertainty:}} Experimental uncertainty must be carried through the calculations to determine the uncertainties in the results. The experimental uncertainties $\sigma \psi_{i}$ are listed in the catalog with the polarization position angles $\psi_{i}.$  One interprets the combination $\{\psi_{i}, \sigma \psi_{i}\}$ for the $i^{\mathrm{th}} $ source to mean that the actual polarization position angle may differ from $\psi_{i}$ by some value  $\delta \psi_{i}$ with the likelihood of  $\delta \psi_{i}$ fitting a normal distribution about $\psi_{i}$ with half-width $\sigma \psi.$ 
	
	To carry the uncertainty in $\psi_{i}$ through the calculations to the results, the Hub Tests were repeated for a total of $10\,000$ uncertainty runs. Each uncertainty run has its own set of polarization position angles $\psi_{i} + \delta \psi_{i},$ where the $10\,000$ values of  $\psi_{i} + \delta \psi_{i}$ for fixed $i$ fit a  normal distribution about $\psi_{i}$ with half-width $\sigma \psi.$
	
\begin{figure}[ht]  
\centering
\vspace{0cm}
\hspace{0in}\includegraphics[bb=130 10 280 185, height=2.3in,keepaspectratio]{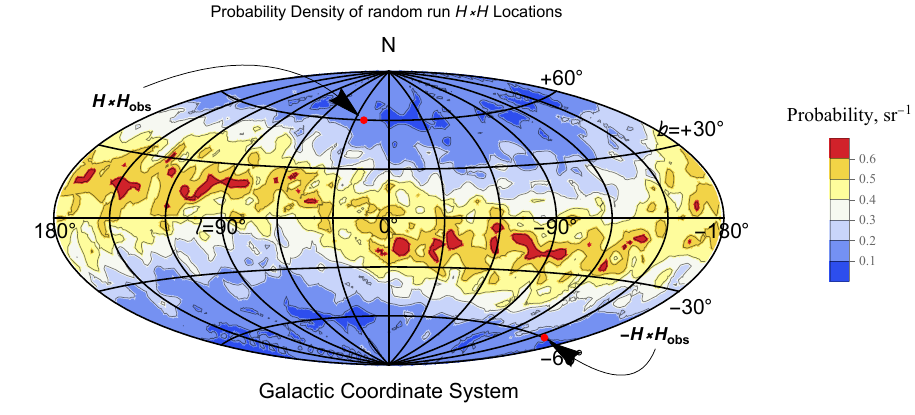}
\caption{(Color online) {{\textit{Probability density of $H \times H$ axis location.} By Figs. \ref{ProbAlign} and \ref{ProbAvoid}, the locations of  $H_{\mathrm{Align}}$ and $H_{\mathrm{Avoid}}$ near the plane of the Galactic Disk are unlikely. By this map, the cross product $H \times H$  is also located in an unlikely region just $30^{\circ}$ from the Galactic Pole.  }  
}}
\label{ProbHxH}
\end{figure}

	For example, each uncertainty run provides a value of the smallest alignment angle $\bar{\eta}_{\mathrm{min}}.$  Those $10\,000$ values of $\bar{\eta}_{\mathrm{min}}$ are fit quite well by a normal distribution with a mean value of $\eta_{0}^{\mathrm{min}}$ = $39.91^{\circ}$ at the distribution's peak and a half-width $\sigma \eta^{\mathrm{min}}$ = $0.38^{\circ} .$ For the purposes of this article, we report the observed value of $\eta^{\mathrm{min}}$ as either the most likely value from uncertainty runs $\eta_{0}^{\mathrm{min}}$  = $39.91^{\circ}$ or the `best' value of $\bar{\eta}_{\mathrm{min}}^{\mathrm{best}}$ = $39.71^{\circ}$ using the catalog values of $\psi_{i}.$  From the uncertainty runs, we have $\bar{\eta}_{\mathrm{min}}$ = $39.91^{\circ} \pm 0.38^{\circ}.$ That result is reported in Table 1.
	
	{\textbf{Significance:}} The significance $p$ of an observed result is the likelihood that random data would produce a better outcome. For that purpose, we made a total of $25\,000$ random runs. The Hub Tests are repeated with the observed QSOs' locations on the sky, but with random directions in place of the observed polarization directions. The fraction of random runs that have better outcomes than is produced by observed data is the significance $p$ of the observed outcomes. 
	
	 For example, for each random run, the smallest alignment angle $\bar{\eta}_{\mathrm{min}}$ is saved and we get a total of $25\,000$ values of  $\bar{\eta}_{\mathrm{min}}.$ The $25\,000$ values are sorted and the rank of the observed value is noted. Just $225$ of the random runs had smaller values of $\bar{\eta}_{\mathrm{min}}$  than the best observed value,   $\bar{\eta}_{\mathrm{min}}^{\mathrm{random}} \leq$   $\bar{\eta}_{\mathrm{min}}^{\mathrm{obs}}$ = $39.71^{\circ}.$ Thus, the significance $p$ of the observed smallest alignment angle $\bar{\eta}_{\mathrm{min}}^{\mathrm{obs}}$ is, by definition, the fraction of runs that produced better alignment, $p$ = $225/25\,000$ = $0.0090\,.$

	Having found previously the uncertainty of  $\bar{\eta}_{\mathrm{min}}$ due to experimental error, the one sigma range of the significance of $\bar{\eta}_{\mathrm{min}}$ can be found. At one sigma below the most likely value for $\bar{\eta}_{\mathrm{min}},$ one finds the angle  $\bar{\eta}_{\mathrm{min}}$ = $\eta_{0}^{\mathrm{min}} - \sigma \eta^{\mathrm{min}}$ = $39.91^{\circ}-$ $0.38^{\circ}$ =  $39.53^{\circ}.$ Out of $25\,000$ random runs, just 130 have values for  $\bar{\eta}_{\mathrm{min}}$ that are less than  $39.53^{\circ}.$ The rank is 130, so the significance of $\eta_{0}^{\mathrm{min}} - \sigma \eta^{\mathrm{min}}$ is $130/25\,000$ = $0.0052.$ See Table 1.

	The significance calculation contributes uncertainty. To estimate the uncertainty $\sigma p$ in the significance of the alignment angle due to statistics, a total of $10\,000$ subsets were created, each with $10\,000$ random runs selected at random from the $25\,000$ random run total. The uncertainty $\sigma p$ is the standard deviation of the $10\,000$ values for $p$ found with the $10\,000$ subsets. This uncertainty $\sigma p$  due to statistics is much less than the uncertainty $\sigma p$  that is due to the measurement uncertainty $\sigma \psi$ and is dropped. Even though the statistical uncertainty is dropped, Table 1 is based on the $10\,000$ subsets of $10\,000$ random runs each. See Table 1 for the significances of the results of the Hub Tests.

\section{Preferred 3D reference frame} \label{3DFrame}

	The preferred Cartesian coordinate system for this 355 QSO sample is based on the locations of the alignment and avoidance hubs, $\pm H_{\mathrm{Align}}$ and $\pm H_{\mathrm{Avoid}},$ the places where the alignment angle function $\bar{\eta}\left( H \right)$ has its extreme values. Note that, in this section, we conflate hubs as points on the Celestial Sphere and hubs as the radial unit vector from the center of the Celestial Sphere to the points on the sphere. The meaning should be clear from the context.

	
	  The radial directions to the hubs $H_{\mathrm{Align}}$ and $H_{\mathrm{Avoid}}$ are separated by $92.3^{\circ} \pm 2.6^{\circ},$ making them perpendicular within experimental uncertainty. Let the alignment hub  $ H_{\mathrm{Align}}$ determine the $z$-axis and let the plane of the hubs $\pm H_{\mathrm{Align}}$ and $\pm H_{\mathrm{Avoid}}$ make the $xz$-plane, with the avoidance hub a couple of degrees off the positive $x$-axis, and on the axis within experimental uncertainty. Then the $y$-axis is in the direction of the cross product of the two hubs, \linebreak$\hat{y} \parallel $ $ \left( H_{\mathrm{Align}} \times  H_{\mathrm{Avoid}} \right)$ or   $\hat{y} \parallel $  $H \times H,$ for short. 
	
	Thus, aside from permutations of $xyz,$ the samples' polarization directions determine the locations where $\bar{\eta}\left( H \right)$ has its extreme values and that determines a Cartesian coordinate frame, the preferred coordinate system for this sample.  Figure \ref{GlobeFromNGP} is drawn in the preferred Cartesian coordinate frame.
		
	The set of $25\,000$ random runs described in Sec. \ref{355QSOs} provides $25\,000$ pairs of alignment and avoidance hubs. From these $25\,000$ pairs, the probability densities on the sky of each axis $H_{\mathrm{Align}},$ $H_{\mathrm{Avoid}},$ and $H \times H$ can be determined. Maps of the probability densities, Figs. \ref{ProbAlign},\ref{ProbAvoid},\ref{ProbHxH}, make it clear that the probability densities have global structure. Global structure from a randomly directed sample begs an explanation.
	
	Since the polarization directions are replaced by random directions, one can conclude that the patterns in Figs. \ref{ProbAlign},\ref{ProbAvoid},\ref{ProbHxH} are a property solely of the locations of the 355 QSOs. By Figs. \ref{ProbAlign} and \ref{ProbAvoid}, one sees that the hubs $H_{\mathrm{Align}}^{\mathrm{(Random)}}$ and $H_{\mathrm{Avoid}}^{\mathrm{(Random)}}$  are most likely to be found near   $(\ell,b)$ = $(-90^{\circ},60^{\circ})$ in the North and $(\ell,b)$ = $(100^{\circ},-65^{\circ})$ in the South. By Fig. \ref{PineNeedleData}, both places host dense populations of sources.
	
	
		
	These observations suggest an explanation for the patterns in Figs. \ref{ProbAlign},\ref{ProbAvoid},\ref{ProbHxH}. Locating a hub far from the sources requires the collective action of many sources and collective action is unlikely with randomly directed samples. The patterns in  Figs. \ref{ProbAlign} and \ref{ProbAvoid} show that, with a discordant randomly directed sample, the most likely locations for alignment or avoidance hubs are near regions with dense populations. 

	One wonders how significant it is that the preferred reference frame is closely aligned with the Milky Way. The latitudes of the hubs $H_{\mathrm{Align}}$ and  $H_{\mathrm{Avoid}}$ listed in Table 1 are $b$ = $15.2^{\circ} \pm 3.3^{\circ}$ and $b$ = $-23.7^{\circ} \pm 2.3^{\circ},$ respectively. That puts them within $30^{\circ}$ of the Galactic Equator. The third $H \times H$-axis has latitude  $b$ = $62.2^{\circ} \pm 3.9^{\circ}, $ just $30^{\circ}$ from the North Galactic Pole. 
	
	 Since  the direction of $H \times H$ is constructed from the other two axes and not independent, the significance of the orientation of the preferred coordinate system is based on the two axes along $H_{\mathrm{Align}}$ and $H_{\mathrm{Avoid}}.$ Of the $25\,000$ random runs, just $738$ have both hubs $H_{\mathrm{Align}}$ and $H_{\mathrm{Avoid}}$ closer to the Galactic Equator than is observed. Thus, the significance of both hubs being close to the Galactic Equator is $738/25\,000$ = $0.030 \pm 0.004,$ the plus-minus due to the uncertainty in the hubs' locations due to experimental uncertainty in the polarization directions $\psi.$  Hence, only one in every $30$ or so randomly directed samples would have the preferred reference system better aligned with the Milk Way.
		
	Thus, the Galactic Coordinate System is not simply a convenient conventional choice, the Milky Way Galaxy plays an essential role in the data, the analysis, and the results. By saturating the optical sky, the Disk keeps the observable optical QSO populations well away from the Galactic Equator. That influences the locations of the sources, which in turn makes the random run patterns of probability densities for  $H_{\mathrm{Align}}$ and $H_{\mathrm{Avoid}}$ in Figs. \ref{ProbAlign} and \ref{ProbAvoid}. And, it is those probability density maps of random run hubs that show that the preferred coordinate system in the directions $H_{\mathrm{Align}},$ $H_{\mathrm{Avoid}},$ $H \times H$  would be unlikely if the polarization directions were random.

\section{Discussion} \label{Discussion}

	The Hub Tests provide four quantities, the minimum and maximum values, $\bar{\eta}_{\mathrm{min}}$ and  $\bar{\eta}_{\mathrm{max}},$ of the alignment angle function $\bar{\eta }\left( H \right)$ as well as the locations on the sky where those extreme values are attained, the hubs $\pm H_{\mathrm{Align}}$ and $\pm H_{\mathrm{Avoid}}.$ From the significances listed in Table 1, the observed values of all four would be unlikely if the 355 QSOs had random polarization directions.

         \vspace{0.5cm}
		         
	\begin{center}
    \begin{tabular}{| c | c | c | c |}
    \hline
    Quantity & Value & Significance & Criteria \\ \hline
    $\bar{\eta }_{\mathrm{min}}$ &  $39.71^{\circ} \pm 0.38^{\circ}$ & $0.0092 \; \{_{-0.004}^{+0.023}$  & $\bar{\eta }$ closer to $0$  \\ \hline
    $\bar{\eta }_{\mathrm{max}}$ &  $49.47^{\circ} \pm 0.38^{\circ}$ & $0.053 \; \{_{-0.015}^{+0.106}$   & $\bar{\eta }$ closer to $90^{\circ}$    \\ \hline
    $H_{\mathrm{Align}}$: $(\ell,b)$ & $(-99.6^{\circ}, 15.2^{\circ}) \pm $ $(3.5^{\circ}, 3.3^{\circ})  $& $0.030 \; \{_{-0.004}^{+0.002}$  & both $H_{\mathrm{Align}}$ \& $H_{\mathrm{Avoid}}$    \\ \hline       
        $H_{\mathrm{Avoid}}$: $(\ell,b)$ & $(-12.3^{\circ}, -23.7^{\circ}) \pm $ $(1.9^{\circ}, 2.3^{\circ})  $ &     & are closer to Equator  \\ \hline
          \end{tabular}
         \end{center}
	

\noindent{Table 1. {{\it{Results of the Hub Tests.}} The angles  $\bar{\eta }_{\mathrm{min}}$ and $\bar{\eta }_{\mathrm{max}}$  determine how well correlated the polarization directions are with respect to one another. The hubs $H_{\mathrm{Align}}$ and $H_{\mathrm{Avoid}}$ are the locations of Virtual North Poles for the Hub Tests of Alignment and Avoidance. The significances are all close to qualifying or qualify as `significant', $p \leq$ $0.05,$ with the significance of $\bar{\eta }_{\mathrm{min}}$ being `very significant', $p \leq$ $0.01.$} 

         \vspace{0.5cm}

		The minimum value of the alignment angle function $\bar{\eta}_{\mathrm{min}}$ shows that the polarization directions act as Local North directions with a significance of about $1\%$ when the Virtual North Pole is placed at the hub $H_{\mathrm{Align}}.$ The maximum value $\bar{\eta}_{\mathrm{max}}$ implies that the polarization directions act as Local Easts with $p \approx$ $5\%$ when the Virtual North Pole is located at the hub $H_{\mathrm{Avoid}}.$ 
		
		The locations of the hubs $\pm H_{\mathrm{Align}}$ and $\pm H_{\mathrm{Avoid}}$ form a preferred Cartesian coordinate system in an orientation that is unlikely to occur by chance. The significance of the orientation is $p \approx$ $3\%.$ 
		
		While the most significant outcome has $p \approx$ $1\%$ for the polarization directions acting as Local Norths, the combined effect of all the results is even more significant. Combining the significances of the Hub Tests' results is complicated because the results may not be independent of one another. Nevertheless, combining the results of the observed polarization directions should put their significance well under $1\%,$  $p < $ $0.01\,.$ Such a correlation is remarkable for a sample that covers a large portion of the sky and has sources at large redshifts. 
		
		\par\noindent\rule{\textwidth}{0.5pt}

		\pagebreak
	
\renewcommand{\theequation}{A.\arabic{equation}}
\setcounter{equation}{0}

\renewcommand{\thefigure}{A.\arabic{figure}}
\setcounter{figure}{0}
	
	\begin{appendix}

	\begin{figure}[ht]  
\centering
\vspace{0.5cm}
\hspace{0in}\includegraphics[bb=50 10 200 185, height=3.in,keepaspectratio]{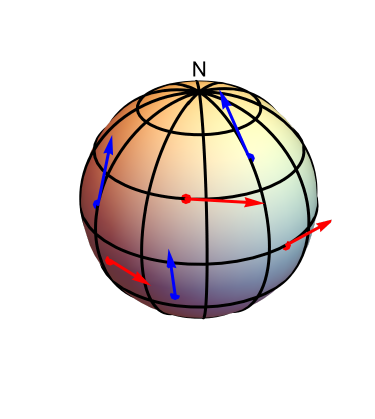}
\caption{(Color online) {\it{Local Norths and Local Easts, alignment and avoidance.}} The blue Local North vectors all `align with a point on the sphere,' the North Pole.  The Local East vectors are tangent to parallels and are perpendicular to the Local Norths. The red Local East vectors `avoid' the direction to the point at the Pole. The Hub Test of Alignment would find the three Local Norths are perfectly aligned with the North Pole, while the Hub Test of Avoidance would find that the Local Easts perfectly avoid the North Pole. }
\label{LATlong}
\end{figure}

\begin{figure}[ht]  
\centering
\vspace{1.5cm}
\hspace{0in}\includegraphics[bb=120 10 270 185, height=2.5in,keepaspectratio]{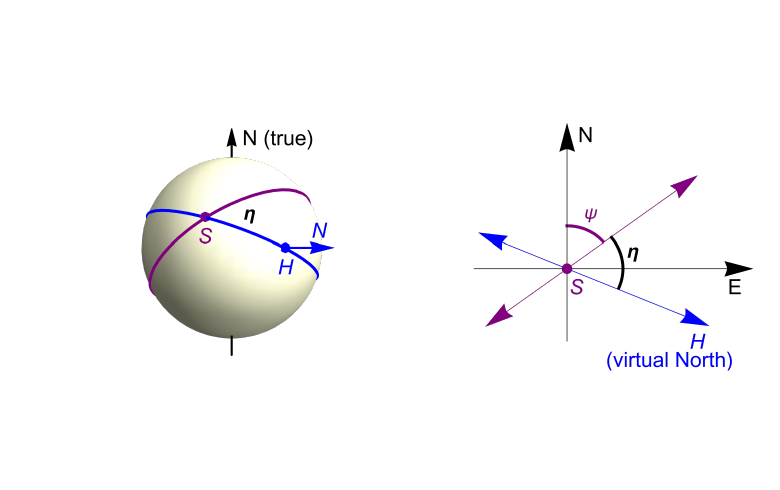}
\caption{(Color online) {{\textit{Alignment angle $\eta$ of the polarization direction with respect to the direction to $H$.}} Left: The Celestial sphere is pictured on the left with true North and a virtual North Pole at an arbitrary point $H.$ The polarization direction determines the purple great circle drawn on the sphere. The blue orthodrome is the virtual meridian from  the source $S$ to the virtual North $H.$  Right: The plane tangent to the sphere at the source $S$. The observed polarization position angle $\psi$ is measured with respect to Local North. Similarly, the `alignment angle' $\eta $ is measured with respect to the virtual Local North toward $H$.   }}
\label{etaH}
\end{figure}

\section{The Hub Test} \label{TheHubTest}

	The Hub Tests are based on notions familiar from the concepts of Local North and Local East directions. At a point $S$ on the sky, the Local North direction is tangent to the great circle `meridian' containing both $S$ and the North Pole. See Fig. \ref{LATlong}. We say that the Local North direction `aligns' with the North Pole. The Local East direction at $S$ is the direction that most `avoids' the North Pole by being perpendicular to the Local North direction. The Local East direction is tangent to a small circle at constant latitude called a `parallel'. Local North and Local East vectors are defined everywhere on the sky, except at the North and South Poles where the definitions fail.
	
	The Hub Tests make every point $H$ on the sky into a Virtual Local North, one point at a time. Fig. \ref{etaH} shows the fundamental quantity for the Hub Test, the alignment angle $\eta.$ The alignment angle $\eta$ is the angle in the tangent plane at $S$ between the polarization direction at the source $S$ and the direction towards the Virtual North Pole at $H.$ 
	
	Recall that the electromagnetic fields of plane waves oscillate, so they are non-oriented,  bi-directional. A bi-directional polarization vector that deviates from the virtual Local North to $H$ by $150^{\circ}$ on one side, deviates by $30^{\circ}$ on the other side. In that case, we take $\eta$  = $30^{\circ}.$   In general, we can always take the alignment angle $\eta$ to be an acute angle,
\begin{equation} \label{EqA0}
	 0 \leq \eta \leq 90^{\circ} \; ,
\end{equation}
because the polarization direction is non-oriented.

	Perfect alignment with the virtual North Pole $H$ occurs when $\eta $  =  $0^{\circ}$ and the direction from $S$ to $H$ coincides with the polarization direction at $S.$ Perpendicular directions, $\eta $   =  $90^{\circ}$, have maximum ''avoidance'' of the direction to the virtual North Pole $H,$ which means the polarization direction is along the virtual Local East direction. The neutral halfway value, $\eta $   =  $45^{\circ}$,  favors neither alignment nor avoidance of the polarization direction with the direction towards the virtual North Pole $H.$

		With $N$ sources $S_{i}$, $i$  =  1, ..., $N$, there are $N$ alignment angles $\eta _{iH}$ for each point $H.$ The average alignment angle  at $H$ defines a function of position $H$ on the sky,
\begin{equation} \label{EqA1}
	\bar{\eta }\left( H \right)  =  \frac{1}{N} \sum _{i=1}^N  \eta _{iH} \; .
\end{equation}
By Fig. \ref{etaH}, each alignment angle, $\eta _{iH},$ for $S_{i} \neq$ $H,$ can be found by applying the arc cosine to one of the following expressions, 
\begin{equation}	\label{Eq2}
	\cos( \eta _{iH}) \; = \;  \mid \hat{v}_{\psi }\cdot \hat{v}_{H} \mid  \; = \; \mid \left(\hat{r}_{S}\times \hat{v}_{\psi }\right) \cdot \left(\hat{r}_{S}\times\hat{v}_{H}\right) \mid  \; = \;  
\end{equation}
\begin{equation} \label{Eq3}
	 \mid \frac{ \cos {\mathit{b}}_S \cos \psi \sin {\mathit{b}}_{H}  + \cos {\mathit{b}}_H \left[ \sin \left( \ell _H - \ell _S \right) \sin \psi  -  \cos \left(\ell _H-\ell _S\right) \cos \psi  \sin {\mathit{b}}_S \right]        }{   \sqrt{1 - \left(\cos \left(\ell _H-\ell _S\right)  \cos {\mathit{b}}_H  \cos {\mathit{b}}_S  + \sin {\mathit{b}}_{H} \sin {\mathit{b}}_{S} \right)^2}    } \mid \; .
\end{equation}
Here $\ell $ and ${\mathit{b}} $ are  longitude and latitude coordinates. The unit vectors $\hat{v}_{\psi }$, $\hat{v}_{H}$ in the tangent plane at $S_{i}$ point in the polarization direction and in the direction toward $H,$ respectively. See Fig. \ref{etaH}. Since the  $\eta _{iH}$ are acute, it follows that the function $\bar{\eta }(H)$ is acute, $0^{\circ}$ $\leq$ $\bar{\eta }(H)$ $\leq$ $90^{\circ}.$  One can show that the function is symmetric across diameters,  $\bar{\eta }(H)$  = $\bar{\eta }(-H).$ We have 
\begin{equation} \label{Eq4}
	0^{\circ} \leq \bar{\eta }(H) \leq 90^{\circ} \quad {\mathrm{and}} \quad  \bar{\eta }(H)  = \bar{\eta }(-H) \; .
\end{equation}
	
	To justify the cross product expression in Eq. \ref{Eq2}, note that the vectors $\left(\hat{r}_{S}\times \hat{v}_{\psi }\right)$ and $ \left(\hat{r}_{S}\times\hat{v}_H\right)$ are in the tangent plane of  $S_{i}$ perpendicular to $ \hat{v}_{\psi }$ and $\hat{v}_H,$ respectively, and are separated by the same angle $ \eta _{iH}$ that separates $ \hat{v}_{\psi }$ and $\hat{v}_H.$ 

	The Virtual North Pole $H$ must not coincide with any of the sources $S_{i},$ $H \neq$ $S_{i},$ because $\eta_{iH}$ is not uniquely defined when $S$ = $H.$ In practice, we consider $0.001$ radian = $0.057^{\circ}$ to be too close and use $\eta _{iH}$ = $45^{\circ}$ when $H$ is within $0.057^{\circ}$ of $S_{i}.$ The replacement applies to at most one source of many and, therefore, is unlikely to change any of the Hub Test results noticeably. 
	
	By (\ref{Eq4}), the minimum of the average alignment angle function  $\bar{\eta }(H)$  occurs at two antipodal points, the `hubs'  $\pm H_{\mathrm{Align}}.$ The two points are the best places to put Virtual North and South Poles for the polarization directions to align with the hubs. With these hubs, the polarization vectors act most like Local North vectors. Thus, the Hub Test of Alignment returns two independent quantities. We get the  minimum of the  average alignment angle function  $\bar{\eta }(H)$, $\bar{\eta }_{\mathrm{min}}$ as well as the locations $\pm H_{\mathrm{Align}}$ where those extreme values occur. 
	
	Similarly, we define the avoidance hub $H_{\mathrm{Avoid}}$ and its opposite $-H_{\mathrm{Avoid}}$ to be the two points on the sphere where the function $\bar{\eta }(H)$ achieves its maximum value, $\bar{\eta }_{\mathrm{max}}.$ With these hubs as Virtual North/South Poles, the polarization vectors act best to avoid the direction to the poles. Thus, the Hub Test of Avoidance yields two independent quantities, the maximum value of $\bar{\eta }(H)$ and the locations of the avoidance hubs $\pm H_{\mathrm{Avoid}}.$
	
	The hubs $\pm H_{\mathrm{Align}}$ and $\pm H_{\mathrm{Avoid}}$ determine a preferred Cartesian coordinate system as described in Sec. \ref{3DFrame} in the main text.
	
	The Hub Tests produce four quantities, the extreme values $\bar{\eta}_{\mathrm{min}}$ and $\bar{\eta}_{\mathrm{max}}$  of the  average alignment angle function  $\bar{\eta }(H),$ as well as the hubs where the extremes are located,  $\pm H_{\mathrm{Align}}$ and  $\pm H_{\mathrm{Avoid}}.$ Also, a preferred Cartesian coordinate system can be constructed based on the locations of the hubs.

		\end{appendix}

\end{document}